\documentclass[conference]{IEEEtran}
\usepackage[usenames,dvipsnames]{xcolor}
\usepackage{graphicx}
\usepackage{epstopdf}
\usepackage{url}
\usepackage{comment}
\usepackage{grffile}
\usepackage{multirow}
\usepackage{authblk}

\begin{document}

\title{A Robust Eco-Routing Protocol Against Malicious Data in Vehicular Networks}

\author[1]{Pavlos Basaras\thanks{pabasara@uth.gr}}
\author[2]{Leandros Maglaras\thanks{leandros.maglaras@dmu.ac.uk}}
\author[1]{Dimitrios Katsaros\thanks{dkatsar@uth.gr}}
\author[2]{Helge Janicke\thanks{heljanic@dmu.ac.uk}}
\affil[1]{Department of Electrical \& Computer Engineering\\ University of Thessaly, Volos, Greece}
\affil[2]{School of Computer Science and Informatics\\ De Montfort University, Leicester, UK}
\affil[ ]{\textit {\{pabasara, dkatsar\}@inf.uth.gr\\ \{leandros.maglaras, heljanic\}@dmu.ac.uk}}

\renewcommand\Authands{ and }

\maketitle

\begin{abstract}
Vehicular networks have a diverse range of applications that vary from safety, to traffic management and comfort. 
Vehicular communications (VC) can assist in the eco-routing of vehicles in order to reduce the overall mileage and CO2 emissions 
by the exchange of data among vehicle-entities. However, the trustworthiness of these data is crucial as false information can heavily affect
the performance of applications.  Hence, the devising of mechanisms that reassure the integrity of the exchanged data is of utmost importance.
In this article we investigate how tweaked information originating from malicious nodes can affect the performance of a real time eco routing mechanism that 
uses  DSRC communications, namely {\it ErouVe}. We also develop and evaluate defense mechanisms  that exploit vehicular communications in order to filter out tweaked 
data. We prove that our proposed mechanisms can restore the performance of the ErouVe to near its optimal operation and can be used as a basis for protecting 
other similar traffic management systems.
\end{abstract}

\IEEEpeerreviewmaketitle

\section{Introduction}
Intelligent Transportation Systems (ITS) incorporate a communications environment over the wireless medium 
between mobile nodes, e.g. vehicles, infrastructure nodes, e.g. road side units (RSUs), with the aim being to increase road safety~\cite{joerer2014vehicular}, \cite{yang2004vehicle}
traffic efficiency~\cite{mariano2015solution} and reduction of CO$_2$ emissions~\cite{maglaras2013exploiting}~\cite{santamaria2015safety}, hence establishing a safer and greener environment for 
transportation~\cite{barba2012smart}~\cite{tsugawa2010energy}. That is, vehicles and RSUs broadcast messages regarding road conditions, accidents, traffic reports, etc. and hence, 
become part of the Vehicular Ad Hoc Network (VANET). 
 Of particular importance are environmental-friendly mechanisms, including 
the reduction of CO$_2$ emissions and mileage~\footnote{\url{http://www.symantec.com/security_response/publications/threatreport.jsp}}~\cite{souza2014decreasing}, since vehicles not powered by fossil fuels are not soon to disappear, e.g. by fully electrical vehicles.

The evolution of vehicles to 
mobile connected entities with On-Board-Units (OBUs) and Internet access~\cite{remy2011lte4v2x} exposes otherwise legitimate vehicles to potential threats, i.e. infected with malware. Reports 
\footnote{\url{http://www.detroitnews.com/story/business/autos/2015/02/08/report-cars-vulnerable-wireless-hacking/23094215/}}
\footnote{\url{http://www.techhive.com/article/221873/With_Hacking_Music_Can_Take_Control_of_Your_Car.html}} indicate that the infection of vehicles is now,
indeed, a realistic scenario and the involvement of such in VANET protocols can result in catastrophic events. Examples range from injecting false data to disrupt
the vehicular environment, e.g. with false data related to traffic congestion, traffic accidents and road conditions~\cite{garip2015congestion}, to inhibiting communication, e.g by jamming~\cite{Punal_2015}, or to  more 
extreme phenomena such as endangering human lives by taking control of a vehicle~\cite{domingo2009safety}.

In~\cite{maglaras2013exploiting} we proposed an eco-routing protocol, namely {\it ErouVe}, which utilizes vehicle-to-infrastructure (V2I), infrastructure-to-infrastructure (I2I) and  infrastructure-to-vehicle (I2V) communications to provide routing instructions to vehicles for a greener trip towards their destination, i.e. optimizing travel duration and CO$_2$ emissions. However, the original {\it ErouVe} algorithm gives no protection against bogus information originating from infected/infiltrated vehicles and 
identifying potential vulnerabilities in a connected car's communication systems is a key factor for shielding it against rational attacks. As online attacks have become potentially more hazardous and aggressive in recent years,
the development of real time defense mechanisms has been stepped up.

To this end, in the current work we focus on providing an effective defense system against
potential spurious data ``running'' through the system's communication phases, which are aimed at disrupting {\it ErouVe}'s routing decisions. Our experimentation shows
that the proposed defense successfully identified outliers and hence, restored ErouVe to near original instructions, i.e. no bogus data was present.
An important information element in VANET communications is the position of adjacent nodes since most applications rely on them. Functions, such as the geographic
routing on the network layer or the V2X applications, require genuine, accurate and reliable location data
regarding neighbors. As a result, we propose to verify the consistency and plausibility of location-related
data of adjacent nodes that are broadcasted frequently as CAMs or geo-networking beacons.

\section{Related Work}

Inter Vehicle Communications (IVC) support applications that are related to safety~\cite{biswas2006vehicle}, traffic management~\cite{milojevic2014distributed} and infotainment, with most of these applications requiring frequent data exchange among vehicles. In addition to reassuring that packets are delivered on
time, which is crucial for safety applications, mechanisms that ensure accuracy and consistency of the data are required. In order to 
provide a secure environment for vehicular communications we need to consider information security requirements, such as confidentiality, integrity and authentication. Also, QoS is important as applications that deal with the safety of the drivers e.g. intersection collision avoidance or emergency braking, require real time communications and have strong delay constraints. 
There are numerous kinds of attack that may threaten confidentiality, availability and authenticity of data~\cite{maglaras2015}.

Many routing protocols try to establish paths among entities that guarantee fast and reliable communication. During the creation of these routes vehicles exchange information about their position, velocity, direction etc. and a mechanism is used to select those nodes that are optimal for each protocol. In a black hole attack, a malicious node exploits this mechanism, advertising itself as providing the shortest path and attracting most of the traffic its way \cite{bibhu2012performance}.  The attacker can choose to drop the packets or manipulate the data, by sending them to the wrong recipient, for example. As a result, the source and the destination nodes
become unable to communicate with each other. Denial of Service (DOS) and Distributed DOS attacks can affect the availability of the data, since the attacker can jam the medium, thereby disrupting the communication among the nodes. The authors in \cite{Punal_2015} showed that RF jamming poses a serious threat to safety in VANETs, for according to their experimental study, jammers can severely disrupt communication up to 465$m$ despite very short communication distances between legitimate devices. 
During a Sybil attack~\cite{sumra2011classes}, a malicious vehicle may pretend to be multiple vehicles  and then use these multiple IDs to distribute
false information. The deleterious effects of such attacks can cascade through the network and cause problems in proper dissemination
of the information. Timing and node impersonation are two other examples of attacks affecting the correct delivery of the information that can be easily launched in a vehicular environment.

A first step towards devising an appropriate defense system is the ability to detect infiltrated vehicles. As noted in~\cite{khan2015detailed}, misbehavior detection in VANETs can be divided into {\it Node-centric} or {\it  Data-centric} mechanisms, with the first inspecting the behavior of a vehicle node, but not the data it sends. For example, 
if the rate at which a node sends packets exceeds a normal (predefined-historical) one, it is characterized as a misbehaving vehicle \cite{maglaras2015}. Other mechanisms in the same category include 
some form of reputation management, which inspects the past and present behavior of a node to derive the probability of future misbehaviour, as implemented in~\cite{kim2012misbehavior}. 

Filtering out false data is another technique widely used in WSNs and VANETs~\cite{cao2008proof}. 
Our proposed scheme is based on a form of reputation and filtering, since vehicles constantly exchange their current information, 
which they use in order to create and maintain a list of their neighbors. In our defense mechanism, all the data collected from the vehicles are gathered and validated by the RSUs~\footnote{\url{http://www.bmvi.de/SharedDocs/EN/Anlagen/VerkehrUndMobilitaet/Strasse/cooperative-its-corridor.pdf?__blob=publicationFile}}.
This way, information that is sent from infected vehicles is discarded and hence, their credibility is considered to be zero.

The second discrimination concentrates on the disseminated data in order to detect misbehaving vehicles, a scheme which is also used in our proposed defense system. Specifically,
the disseminated data are evaluated for {\it plausibility} and/or {\it consistency}. For example in our evaluation scenario, plausibility will ensue if a vehicle reports
a travel time of a few seconds while traveling a relatively long path. Consistency will be applied if a vehicle sends high (or low) statistics for a road segment, e.g. CO$_2$
emissions depending on the attack's goal, which although plausible, significantly deviate from similar reports of
vehicles from their one hop neighborhood.

\section{Preliminary Work, $ErouVe$}
\label{erouve-pre}

The original $ErouVe$ algorithm, as presented in~\cite{maglaras2013exploiting}, identified congestion phenomena by taking into consideration 
the travel duration and CO$_2$ emitted by vehicles in specific road segments. In the next subsection we describe the algorithm
specifications and functionality along with the new mechanism for routing instructions.

\subsection{System Description}
\begin{figure}[!htb]
  \centering
  \includegraphics[width=\linewidth]{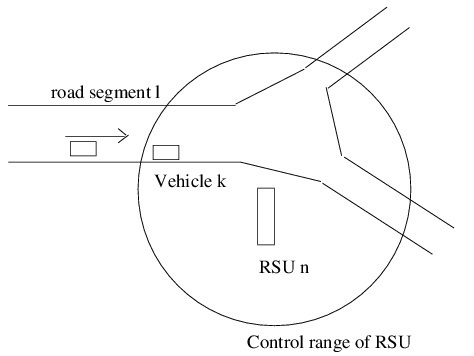}
  \caption{Decentralized CO2 reduction system based on DSRC communications}
  \label{toy_net}

\end{figure}

We consider a network system $G=(V,L)$, where $V$ depicts the set of nodes (intersections - RSU placements) and $L$ are the road segments connecting those intersections. The set of road segments adjacent to an RSU $n\in V$, is denoted as $S(n)$. RSUs with
common adjacent road segments are considered as neighbors, e.g. of $n$, and denoted as $N(n)$. Note that two neighboring RSUs may be connected through
more than one route. Vehicles send data regarding their traversed road segment $l\in L$, i.e travel duration and CO$_2$ emissions, to the corresponding RSU (Figure~\ref{toy_net}). Next, 
neighboring RSUs exchange beacon messages with the data acquired from vehicles and with these specifications, each RSU $n$ calculates average values for each segment 
$l \in S(n)$. In order to have updated information for a road segment, the RSUs only consider records within the most recent time window of $s$ seconds (TIN), from which 
an optimal-eco route for each vehicle can be identified.  
Note that ErouVe runs on level 2 of automation to advise upcoming vehicles;  "Combined function automation".

\subsection{System Initialization}
The initial step of the system is to compute for all $n\in V$ their corresponding neighbors, i.e. $N(n)$ and for all $m \in N(n)$, Dijkstra's algorithm is used 
to acquire the distances between two RSUs,  $D_{nm}$, based on GPS data. 
Consequently, each RSU $n$ becomes aware of it's vicinity and the road segments through which it is connected to any other RSU $m\in N(n)$. Note that no time or CO$_2$ cost is initially calculated for the road segments. Table~\ref{rsu_table} briefly describes
the initial information stored by each RSU. As illustrated, column $2$ holds the neighbors of each RSU, column $3$  has the road segment(s) through which neighboring RSUs are
connected and finally, column $4$ illustrates the distances of each road segment.  For example, a vehicle $k$ from R1 can reach R2 through segments $l_a$ and $l_b$ in distances D$_{a}$ and D$_{b}$, respectively.

\begin{table}[h]
\centering
\caption{Example of Connections Table for $3$ RSUs}
\begin{tabular}{|l|l|l|l|}
\hline
RSU\_Id & Neighbors    & Road Segments                                                              & Distance                                                                                              \\ \hline
R$_1$   & R$_2$, R$_3$ & \begin{tabular}[c]{@{}l@{}}R$_2$: l$_a$, l$_b$\\ R$_3$: l$_c$\end{tabular} & \begin{tabular}[c]{@{}l@{}}R$_2$: l$_a$(D$_{a}$), l$_b$(D$_{b}$)\\ R$_3$: l$_c$(D$_{c}$)\end{tabular} \\ \hline
R$_2$   & R$_1$, R$_4$ & \begin{tabular}[c]{@{}l@{}}R$_1$: l$_a$\\ R$_4$: l$_d$\end{tabular}        & \begin{tabular}[c]{@{}l@{}}R$_1$: l$_a$(D$_{a}$)\\ R$_4$: l$_d$(D$_{d}$)\end{tabular}                 \\ \hline
R$_3$   & R$_1$, R$_5$ & \begin{tabular}[c]{@{}l@{}}R$_1$: l$_b$\\ R$_5$: l$_e$\end{tabular}        & \begin{tabular}[c]{@{}l@{}}R$_1$: l$_b$(D$_{b}$)\\ R$_5$: l$_e$(D$_{e}$)\end{tabular}                 \\ \hline
\end{tabular}
\label{rsu_table}
\end{table}

\subsection{Communication Phases}
This section briefly explains the different communication phases of the original algorithm. 

\subsubsection{{\bf Road Segment Measurements (I2V)}}
For any vehicle $k$, which just completed its course on road segment $l$  the corresponding RSU impels vehicle $k$ to:

\begin{itemize}
\item calculate total time traveled ($TT_{lk}$), and CO$_2$ emissions ($C_{lk}$) on road segment $l$. 
\item send to the RSU the calculated values of  $TT_{lk}$ and $C_{lk}$
\end{itemize}

\subsubsection{{\bf Communication of RSUs (I2I)}}
Each RSU will send the accumulated values for mean travel time and CO$_2$ emissions of each vehicle to the corresponding neighboring RSUs through beacon messages.

\subsubsection{{\bf Route Request-Reply (V2I)-(I2V)}}
Each vehicle $k$ that enters the control range (intersection area) of an RSU sends a route request message ($R_q$) to the corresponding RSU,
which in turn, after solving the optimization problem (cf. next subsection) based on data obtained through I2I,  sends routing instructions
 to the corresponding vehicle via an $R_a$ message  (route answer).

\subsection{New Decision System for Optimal Routes}
\begin{figure}[!htb]
  \centering
 \includegraphics[width=\linewidth]{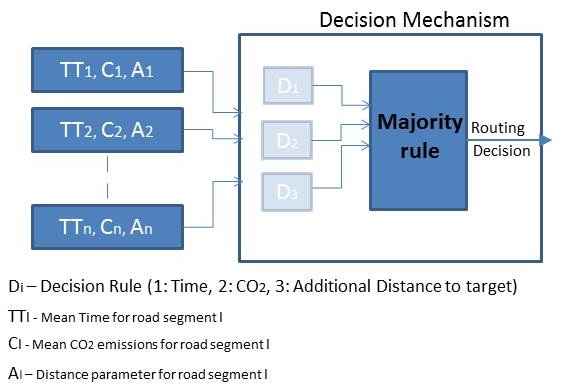}	
  \caption{New decision mechanism}
  \label{dec-syst}
\end{figure}

In the initial ErouVe mechanism, as presented in \cite{maglaras2013exploiting}, weights were assigned to each segment adjacent to the current road and then 
the road with the minimum was chosen.
By following  a slightly different approach we developed a multiple decision mechanism. The new mechanism, rather than adding the different values of the three features used, e.g. Time, CO2 and distance, it logically combines the outcomes of the three decision rules, each representing one of them (Figure~\ref{dec-syst}). 

In the new ErouVe mechanism, the RSU, after receiving a route request message from an approaching vehicle $k$, compares the outgoing road segments based on the current mean time, mean CO2 and the added distance that each routing decision brings about. The outcomes of each decision are combined using weighted majority  voting and different weights can be used in order to focus on one of the different optimization parts, e.g. time, distance or CO2
emissions. In the default system settings, all optimization parts have the same significance. For example when comparing two potential routes, e.g. $k$ and $l$, if D$1$ and D$2$ for $k$ are greater than D$1$ and D$2$ respectively of $l$, $l$ is selected as the next road segment.

\section{ErouVe Vulnerabilities}
\label{erouve-vulnerabilitues}
As previously noted, the original {\it ErouVe} algorithms utilize V2I, I2I and I2V communications, in order to ascertain which is the most eco-friendly route
for any vehicle to follow. However, the technique's performance so far, assumes that vehicles will send only {\it real} data to the corresponding
RSU. If we devise a scenario where {\it tweaked} information exists among the received data, the algorithm's formula can mislead vehicles to not only
false eco-friendly routes, but also, create traffic congestion and hence, significantly deteriorate the system's performance, i.e. increase travel time and 
CO$_2$ emissions.

In this study, we classify tweaked information into two basic categories depending on how an infiltrated vehicle manipulates data: 

\begin{itemize}
\item Send tweaked data to {\it favor} a route (FAV)
\item Send tweaked data to {\it fend} from a route (FEN)
\end{itemize}

FAV can be seen as an attack that creates a false image for a specific road segment, by sending relatively 
small statistics, i.e. short travel time or CO$_2$, thus making a target route favorable. In such a case, vehicles could be instructed to follow the {\it attacked} 
route,  however, if the road throughput cannot satisfy the increasing number of vehicles, this can result to traffic congestion and bottlenecks. 
FEV also tweaks the real conditions regarding the road segment under consideration, but follows a reverse policy from FAV, e.g. sending a relatively 
large travel time to the corresponding RSU. With such misinformation, vehicles will be directed to a different path which can also result in the aforementioned 
problematic scenarios.

However, modified data regarding the accumulated CO$_2$ emissions or travel time is not the only vulnerability of the original {\it ErouVe} algorithm.  Recall 
that once a vehicle exits the road segment under consideration, it sends a report  to the corresponding RSU about the ``condition'' of the road segment it has traversed.
However, so far RSUs have had no knowledge of which route the corresponding vehicle actually followed, apart to what was stated by the sending vehicle itself,
and thus, cannot distinguish to which route the received data belongs.  Consequently, an infiltrated vehicle can denote that these values correspond to a different route 
(regardless of whether these  values are altered or not) and hence, meddle with the system's next decisions.
With the above considerations, the original algorithm stands unprotected (vulnerable) to such false information and thus, our primary objective lies in devising a defense system
to counter data originating from such malicious vehicles.

\section{Attack Plans}
\label{attack-plan}

\subsection{Attack Objectives}

To built on our defense system, we discuss several attack plans and their impact on {\it ErouVe}. The original {\it ErouVe} algorithm
was implemented in order to balance the traffic flow between all possible available routes with a common destination and hence, solve 
potential road congestion. The proposed technique  was compared to a scenario where the shortest route, followed by all vehicles, 
was unable to satisfy the traffic flow, thereby creating congestion in the path. By experimenting in high density 
traffic conditions, we found that {\it ErouVe}'s routing instructions successfully managed the traffic flow between the 
corresponding available paths and as a consequence, significantly enhanced the system's performance, i.e. up to 30\% improvement in travel duration. 
As a result, our attack plan focuses on sending ``appropriated'' (tweaked) data to recreate a scenario where all vehicles follow the shortest
path and create congestion, although under the {\it ErouVe} paradigm. Intuitively, a combination of attacks, i.e. vehicles 
sending {\it favorable} statistics regarding the shortest road segment, i.e.  FAV, and complementary {\it unfavorable} ones for the other route(s), i.e.  FEV, 
will affect the systems routing decisions. By reversing the attack plan on the road segments, i.e. FAV for the longer routes and FEV for the shortest path, 
we obtain a different impact on the protocol's routing decisions. In this case scenario, vehicles will unnecessarily be rerouted to longer routes, resulting 
in increased travel duration and CO$_2$ emissions for each individual vehicle and concurrently, the system.

The aforementioned attack plans have contradictory objectives. In the current study, we focus on the recreation of congestion for the shortest route by 
exploiting the vulnerabilities of the original protocol, i.e. Fake Route (FR) and Fake Data (FD).





\subsection{How To Attack}
First, recall that {\it ErouVe} uses data collected from vehicle measurements, accumulated within the most recent time window of $s$ seconds,
i.e. in TIN and hence, bogus information has a maximum lifetime of TIN in {\it ErouVe}. Moreover, our experimentation showed that data from a 
single infiltrated vehicle can have zero effect in the original {\it ErouVe} protocol, i.e. does not sufficiently change the weight values assigned to road
segments and thus their overall ranking, depending on the extent to which the data are tweaked from their original values. However, if an attacker tries to use significantly
deviated values to affect the formula/protocol, the received data from other (healthy) vehicles in a relatively short time, would render the identification of such bogus vehicles an easy task.

Since a single bogus vehicle may not make a difference to the protocol's routing decisions, grouped attacks are necessary, i.e. a number of infiltrated cars that 
report their stats to an RSU for a target road segment in a relatively short time. However, bogus information has a lifetime
TIN in {\it ErouVe} and thus, short time reports must be defined with respect to TIN.  As a final observation, on the occasion 
where a successful attack occurs, the system can still recover quickly if the weighted order of road segments is not changed much and a sufficient number of healthy 
(non tweaked) vehicle reports follow. Consequently, catastrophic results, i.e. creating traffic congestion or unnecessarily rerouting a large of number vehicles to longer routes, 
can still be avoided, even with no sophisticated protection against false information.

To summarize, vehicles must not only meddle with the data to a degree that will not be undone with a few upcoming healthy vehicles, but also, to such an extent 
that it will not make the RSU suspicious, i.e. it cannot send extremely deviated values from the actual measurements. Finally, timed attacks are essential
with respect to TIN as a single vehicle might not make a difference in the overall ranking of the road segments.

\section{Proposed Defense System: Enhanced ErouVe}
\label{prop-defense}

The goal of our defense system is to filter out tweaked data, so as to return the functionality of {\it ErouVe} to near identical routing decisions, i.e. to an attack
free scenario. Hence, data received by an RSU will be ``judged'' for both {\it plausibility} and {\it consistency}~\cite{khan2015detailed}. 

\subsection{Fake Route Countermeasures}

In order to counter the fake route problem we utilize the yet unused communication phase, i.e.  {\it Vehicle-to-Vehicle} (V2V) communication in our model.
 To this end, vehicles traveling for instance on a specific road segment $l$, broadcast beacon messages regarding the vehicle's ID and that of their 
current road segment, e.g.  $l$. Upon exiting the road segment under consideration, a vehicle $k$ 
now sends information regarding, not only $TT_{lk}$ and $C_{lk}$, but also, the vehicle IDs that co-traveled with vehicle $k$ on road segment $l$. 

By instructing vehicles to gather information about their vicinity in their current road segment, bogus vehicles cannot state a different route than the actual one 
they followed. This is due to the fact that the current mechanism allows an RSU to have an accurate image for which vehicle followed which route based on the majority 
of votes. To bypass the system's new defense,  a large number of infiltrated vehicles need to be grouped appropriately, i.e. of magnitude greater than
the currently healthy vehicles in the corresponding road segment. Nonetheless, in such a scenario, where the majority of vehicles are infected vehicles,
all defense mechanisms are bound to fail.  In our experimentation, we assume that beacons exchanged between vehicles cannot be ``heard'' in different road 
segments. This can be justified if we consider that the distance between the road segments could be greater than the standard DSRC communication range or because of the existence of obstacles, e.g. buildings in an urban scenario that interfere with the communication.

%

\subsection{Fake Data Countermeasures}

After properly matching data to the corresponding routes, we have to deal with vehicles that tweak their accumulated statistics of travel duration 
and CO2 emissions. First, we assume that statistics from healthy vehicles in short time, e.g. of a few seconds, cannot deviate significantly. It is a reasonable assumption 
if we consider that nearby vehicles will experience similar traffic conditions, e.g. similar traffic density. Now, we need to clarify the validity of each newly received vehicle report.
To this end, we define a new time window of about a third of TIN, namely, {\it Validation Window (VoW)}, which will hold the reports of vehicles in a very recent image of  
the road segment under consideration. The Euclidean Distance between the report under ``judgment'' and those in VoW will decide the validity of the new data:
\begin{equation}
D(x) = \sqrt {\sum_{i=1}^{N}(x-y_i)^2}
\end{equation}
where, $x$ stands for  CO2 emissions (or travel duration) of the new vehicle and $y_i$ for the corresponding $N$ values in VoW.
$D_{x}$ is compared to a threshold ($TH_{d}$) that determines to which set it will be included based on the rule:
If $D_{x} < Th_{d}$ then x $\in$ VoW else x $\in$ PBS. Parameter $TH_{d}$ determines the sensitivity of the defense mechanism when categorizing new data as normal or bogus, cf. subsection~\ref{eval-params}.

\begin{figure}[!htb]
  \centering
     \includegraphics[width=\linewidth]{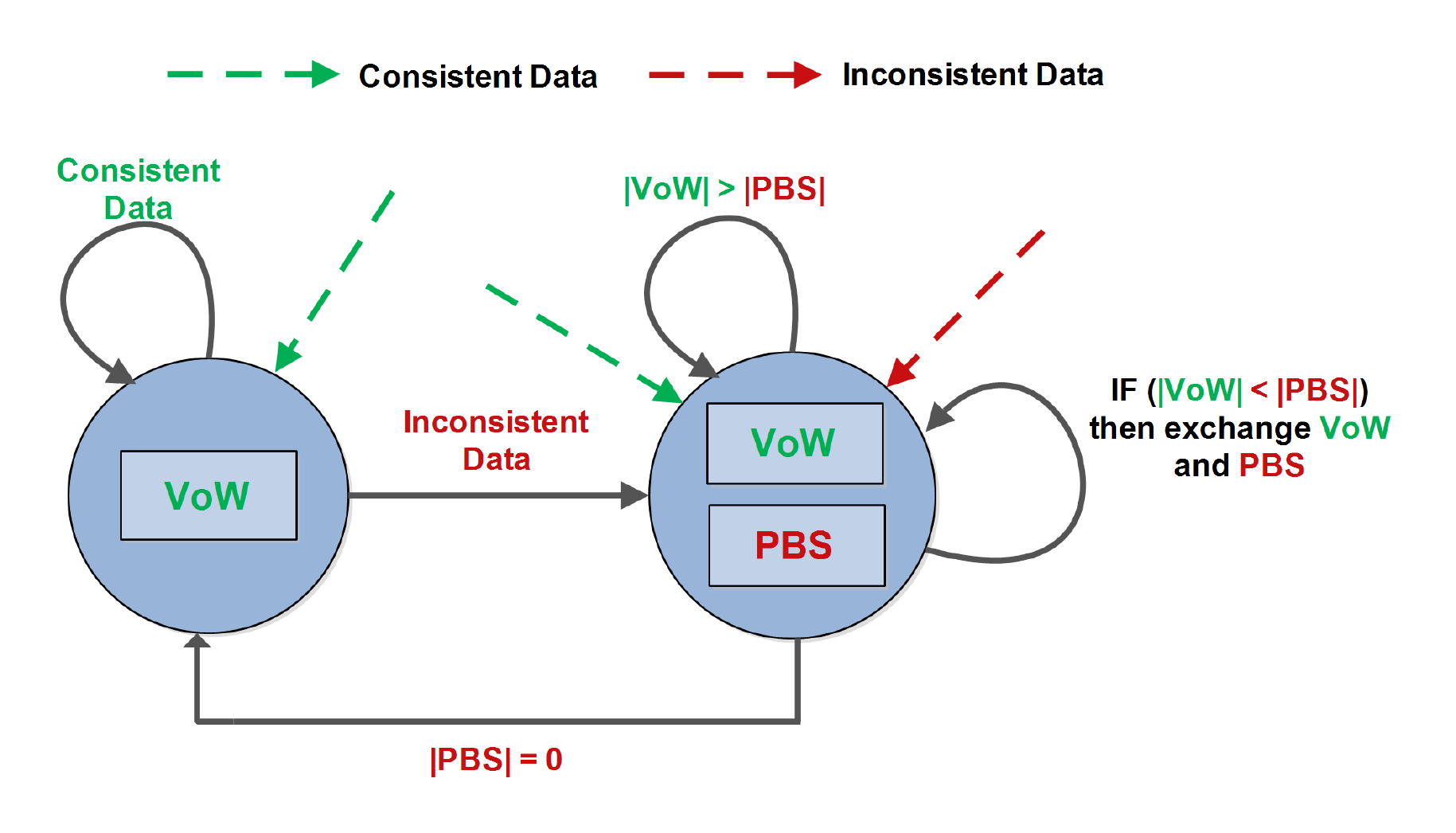}
  \caption{Fake Data Countermeasures}
  \label{shift-lists}
\end{figure}

However, a {\it distant} report is not necessarily a bogus one, i.e. it may correspond to a true change in the traffic conditions of a road segment from dense
to light traffic (congested to uncongested) and vice versa. Consequently, once a distant vehicle is identified, we do not take prompt action to drop its data, but rather save them in a separate set, namely, {\it Potentially Bogus Set (PBS)}
in order to account for the abovementioned case. We expect that if the report corresponds to a realistic traffic change, a number of similar ones are to follow. If the upcoming values are consistent with those in VoW, then the values in PBS are dropped and labeled as truly bogus
data. Alternatively, if the size of PBS 
grows beyond that of VoW, we acknowledge a traffic shift and thus, integrate values of  PBS to VoW. Figure~\ref{shift-lists} illustrates the proposed mechanism. Data are consistent (VoW) when below the distance threshold and otherwise inconsistent (PBS).

Finally, we should note that, as explained in Section~\ref{erouve-pre}, a vehicle sends an $R_{q}$ message in order to 
receive instructions. This places the following constraint: vehicles cannot easily lie about their travel duration. This is due to the fact that the RSU
is aware of the time interval between the reception of an $R_{q}$ message, and the time it receives the statistics from the corresponding vehicle. Nonetheless, more sophisticated plans can be deployed to tweak travel duration, but are beyond of the purposes of the current study. Henceforth and without loss of generality we assume that only CO2 emissions are tweaked.

\section{Experimentation Settings}

\subsection{Simulator}
For the evaluation of our model, we use the simulator VEINS~\cite{sommer2011bidirectionally}, which is composed of two well known simulators: OMNET++ an event-based network
simulator and SUMO, a road traffic simulator. To calculate CO$_2$ emissions for each individual vehicle we apply the EMIT model integrated in VEINS.
It is a statistical model  for instantaneous emissions and fuel consumption based on the speed and acceleration of light-duty vehicles. 

\subsection{Evaluation Scenario}
\begin{figure}[b]
  \centering
     \includegraphics[width=\linewidth]{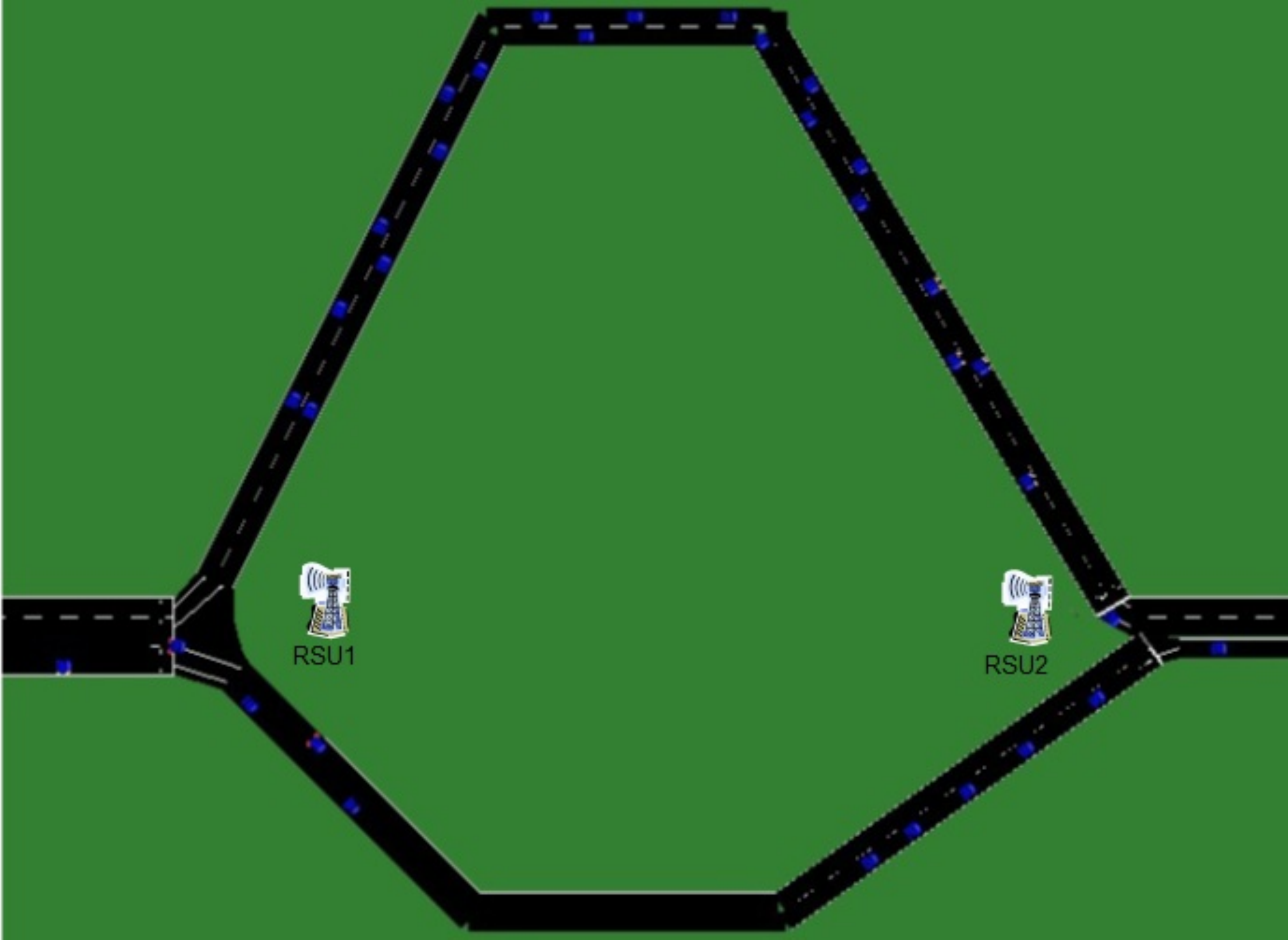}
  \caption{Simulation Map}
  \label{map}
\end{figure}

Similarly to our previous work \cite{maglaras2013exploiting}, we built a map about $2km$ long (Figure~\ref{map}) with a single direction and two available paths. The upper and longer path is about $275m$ long, whereas the lower and shorter path is about $190m$. Both road segments have the same capacity in lanes, i.e. $2$ lanes. These paths merge at junction $2$, where the upper part can occupy $2$ lanes of the next $3$ lane road segment, whereas the lower part can occupy only $1$. This setting is used to demonstrate a typical urban scenario, where part of a road can be temporarily closed due to maintenance or a car accident. Another potential scenario includes crossroads with different priorities, where vehicles in the road segment with less priority line up and give room to traffic flows on roads with higher priority. Such considerations coupled with medium traffic can make a road segment that seems attractive, i.e. shorter path towards destination, unable to satisfy the traffic demand and consequently, result in major traffic congestion. 

\subsection{Communication Settings}

\begin{itemize}
\item {\bf Communication Range}: this is the communication range that can be achieved from vehicles or RSUs according to the setup of the system, which in our experimentation is set to $300m$.
\item {\bf Handshake Range}: this is the range after which an approaching vehicle is aware of the presence of an RSU at an upcoming intersection through beacon
messages emitted by the RSU. At this point, vehicles store the position of the corresponding RSU and this range is set to $100m$.
\item {\bf Control Range}: the final communication range of our system depicts the distance at which vehicles receive routing instructions ($R_a$ message) from an RSU.
In our simulation we set this range to a medium value, in order, if necessary, to give time to vehicles to perform rerouting, i.e. $50m$.
\end{itemize}

\subsection{Parameters}
\label{eval-params}
In Sections~\ref{erouve-vulnerabilitues} and~\ref{attack-plan}, we explained the vulnerabilities of the original ErouVe algorithm and  devised attacks to address those points. Table~\ref{params} summarizes the attack plans and their configuration, vehicle velocity, number of vehicles, and TIN values, as used in our experimentation. Group size is the number of consecutive vehicles that report false data, i.e. one to five vehicles, and attack interval is the time between such groups, e.g. every six seconds. The attack intervals are chosen with respect to TIN, i.e. at least two attacks groups must occur within one TIN. {\it opt} indicates how bogus vehicles tweak their original values in order  to deceive the system. It is calculated for each road segment with respect to the road length and vehicle velocity, i.e assuming vehicles travel in an uncongested road segment with the maximum allowed speed. For the FR attack, vehicles do not tweak their data,
but rather, state that the accumulated statistics correspond only to the long route. For FD, bogus vehicles traversing the short route will say that they have  experienced uncongested road conditions, i.e. opt, whereas for the long route vehicles will state that there is significant congestion. Both attack protocols favor the short route in hopes of creating congestion. Extensive experimentation was conducted in relation to the simulation parameters and in the next section, we present the most characteristic results. Unless stated otherwise, default values are used.

\begin{table}[!htb]
\centering
\caption{Simulation Parameters}
\begin{tabular}{|c|c|c|}
\hline
{\bf Parameters}     & {\bf Range} & {\bf Default} \\ \hline
Attack Type          & FR, FD      & FD            \\ \hline
Group Size           & 1-5         & 3             \\ \hline
Attack Interval (s)  & 6,10,14     & 10            \\ \hline
FR Short Route       & opt-2*opt   & original           \\ \hline
FR Long Route        & opt-2*opt   & original           \\ \hline
FD Short Route       & opt-2*opt   & opt           \\ \hline
FD Short Route       & opt-2*opt   & 2*opt         \\ \hline
Infected Vehicles (\%)   & 10 - 30      &   20           \\ \hline
TH$_d$ (\%) & 10 -  50      & 10           \\ \hline
Vehicle Speed (Km/h) &40 - 90       & 40            \\ \hline
Number of Vehicles   & 50 - 150      & 150           \\ \hline
TIN (s)   & 30 - 120      & 30           \\ \hline
\end{tabular}
\label{params}
\end{table}

\section{Performance Evaluation}
\label{perf-eval}

\subsection{ErouVe Vs Shortest Path VS FR attacks}

\begin{figure}[!htb]
  \centering
    \includegraphics[width=\linewidth]{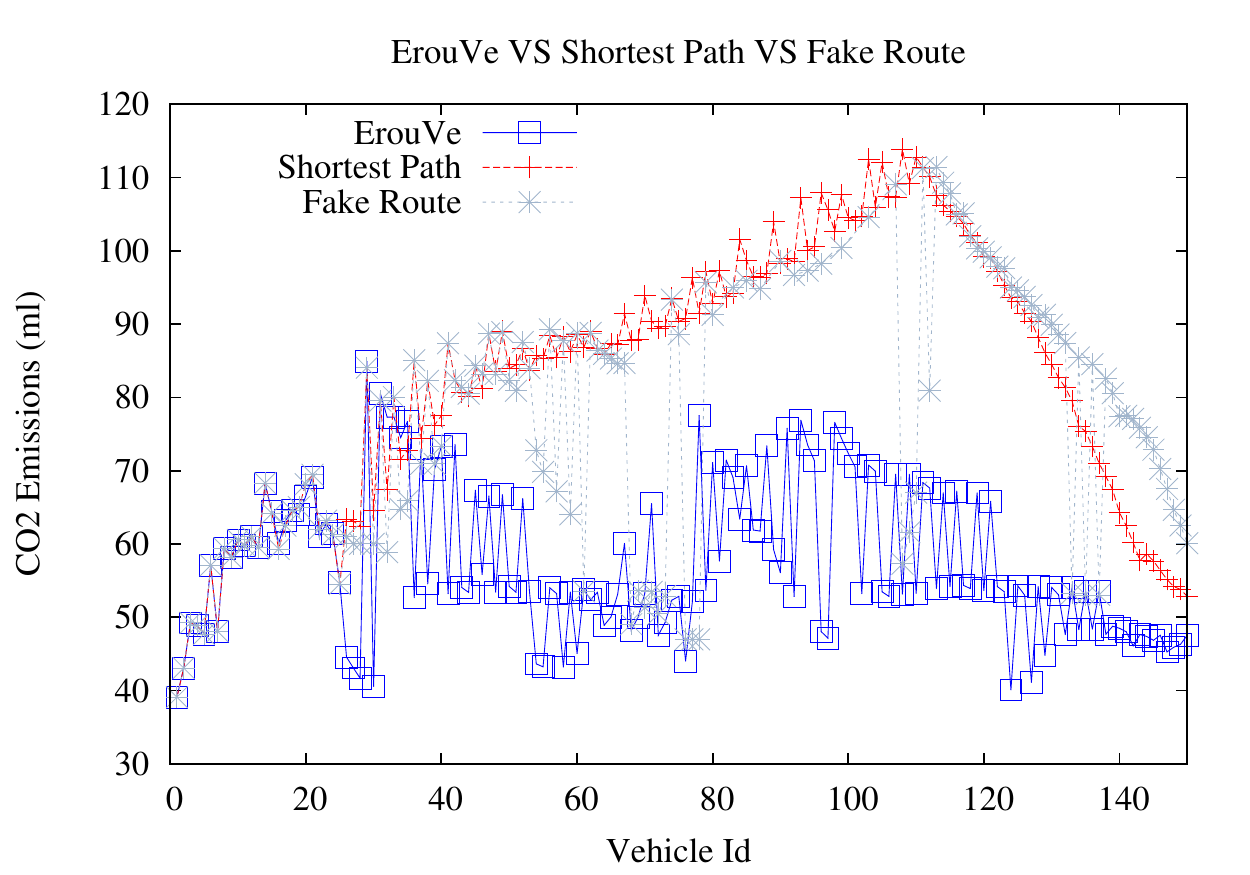}
    \includegraphics[width=\linewidth]{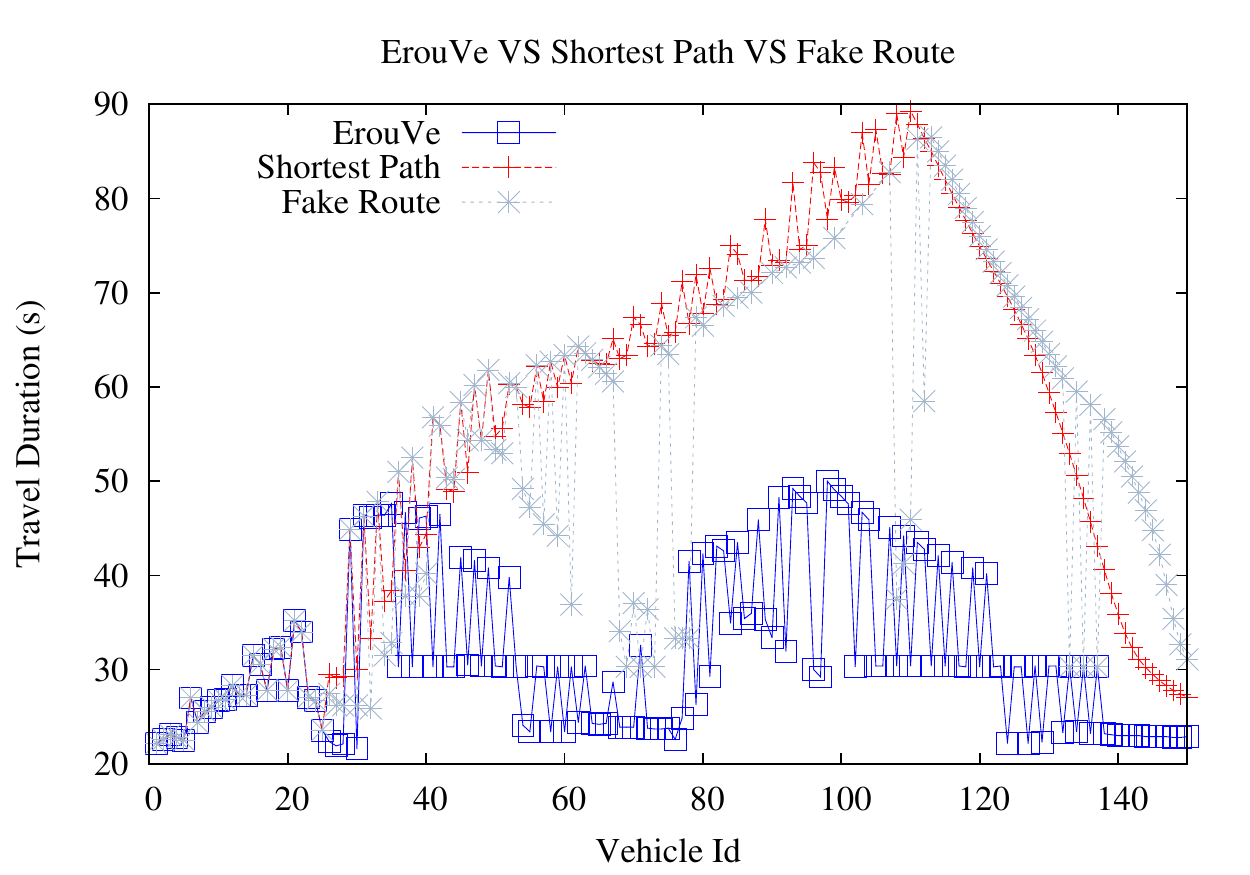}
  \caption{FR successfully deceives the original algorithm into sending vehicles to the short route and thus creating congestion. Travel duration and CO2 emissions are significantly increased by 31\% and 20\% respectively.}
  \label{fake-route}
\end{figure}

In Figure~\ref{fake-route}, the CO2 emissions (ml) and travel time (sec) of each vehicle are demonstrated.
ErouVe in an unprotected mode performs similar to the original shortest path, since due to the
fake route attack it sends most of the vehicles to follow the lower road segment (shortest path).
This increased traffic leads to road congestion that has an immediate effect on both time and CO2 emissions. That is, the mean increases in time and CO2 compared to that in the attack free scenario are 31\% and 20\%, respectively.
Such an increase can be further explained considering that ErouVe sends 25\% of the vehicles to follow the longer route, whereas in the FR scenario only about 8\% of the vehicles take the longer path.
Such observations justify the need for countermeasures and the proposed defense mechanism, 
as described on Section~\ref{prop-defense}, makes the ErouVe mechanism robust to such attacks.

\subsection{Impact of Attack Group Size}

\begin{figure}[!htb]
  \centering
     \includegraphics[width=\linewidth]{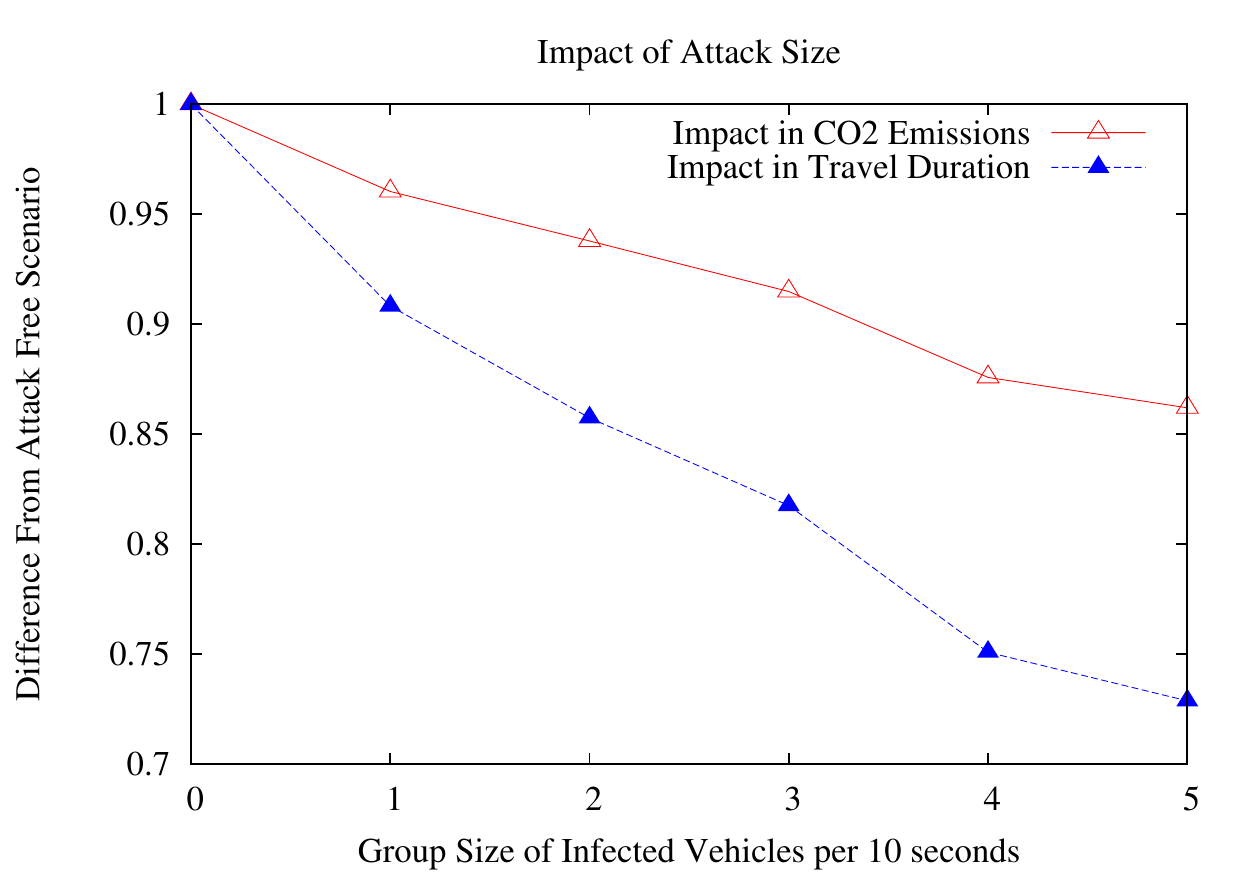}
  \caption{As the number of  FD attacks running in system increases, ErouVe's performance drops. About 30\% of vehicles out of the total simulation were bogus (attack group size set to 5) for a 25\% decrement in travel duration.}
  \label{group-size}
\end{figure}

Figure~\ref{group-size} illustrates how the number of consecutive vehicle attacks (attack group size) affects the system's average performance, with the attack interval set at 10 seconds. The Y-axis represents the deviation from an attack free scenario, i.e. performance drop. For one vehicle per 10 seconds we observe a minor deviation, for example, lower than 5\% in CO2 Emissions. As the attack group increases and thus more bogus data are running the system, the unprotected ErouVe mechanism is further deceived, e.g. more than 25\% increase in travel duration for five vehicles per attack group. It is worth noting that one attacker 
per 10 seconds depicts 8.6\% of 150 vehicles, while for a group of five vehicles, the bogus community rises up to 30\%. Although this observation indicates a strong point for ErouVe, i.e. it takes a large number of vehicles to drop its performance about 25\%, it also
highlights the necessity for a defense mechanism capable of spotting spurious data to ``cure'' the system.

\subsection{Impact of Attack Interval}

\begin{figure}[!htb]
  \centering
     \includegraphics[width=\linewidth]{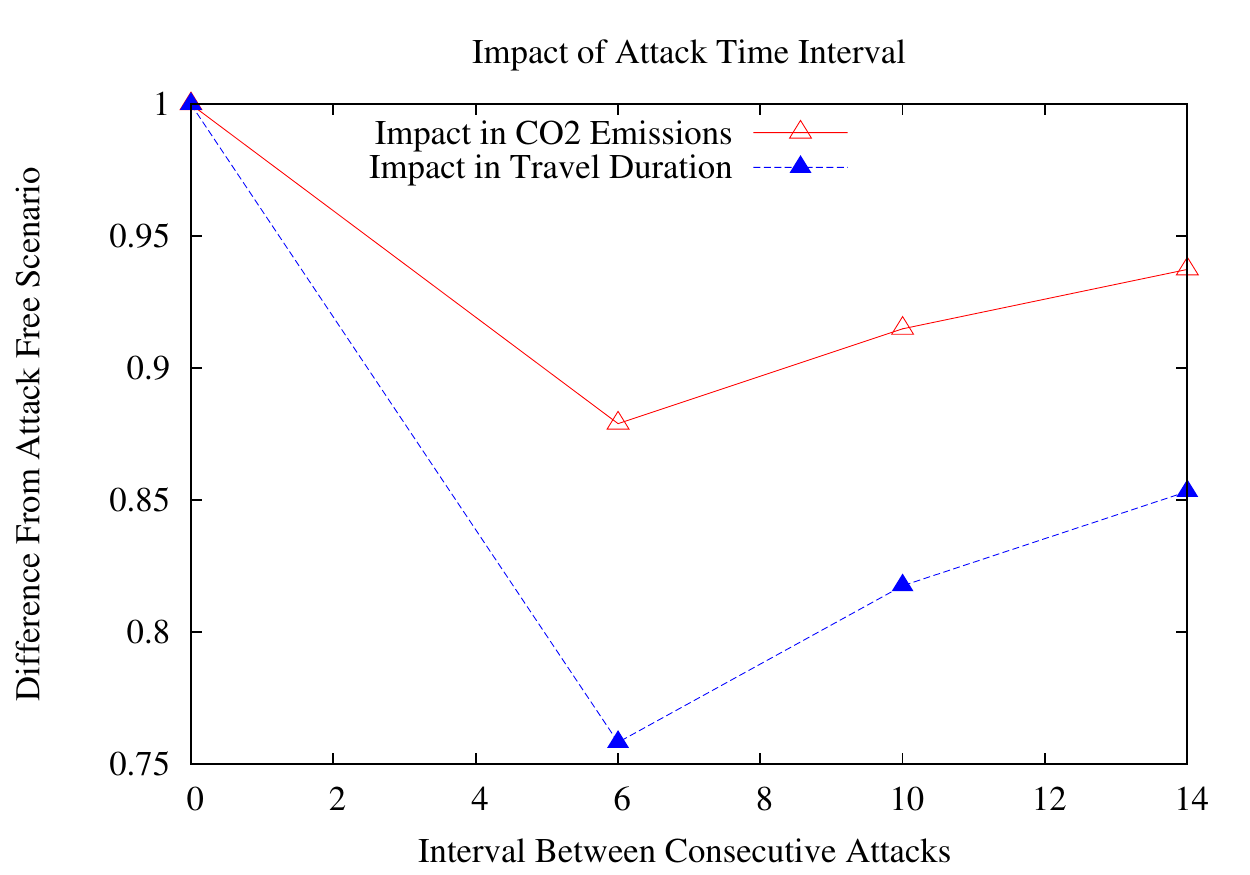}
  \caption{In order to significantly affect the routing decisions of ErouVe, bogus data need to arrive in a timely manner, so as to continuously have bogus data in the system. Otherwise ErouVe may quickly recover to original routing instructions.}
  \label{att-inter}
\end{figure}

In  Figure~\ref{att-inter}, we investigate the frequency of the attacks with the attack group size set to three vehicles. As illustrated, more frequent attacks have greater impact on the performance of ErouVe, e.g. about 24\% in travel duration when attacks happen every six seconds, whereas there is 15\% performance drop when the interval is 14 seconds. Note that for the interval of 14 seconds, only two attack groups ``fit'' in TIN, which explains the lower impact in the protocol's performance. As the simulation time flows, the impact of
earlier bogus data expires and consequently if no significant amounts of new such data are received in a short time, the system is very likely to recover to near normal routing decisions.

\subsection{Impact of Defense System VS FD attacks}

\begin{figure}[!htb]
  \centering
     \includegraphics[width=\linewidth]{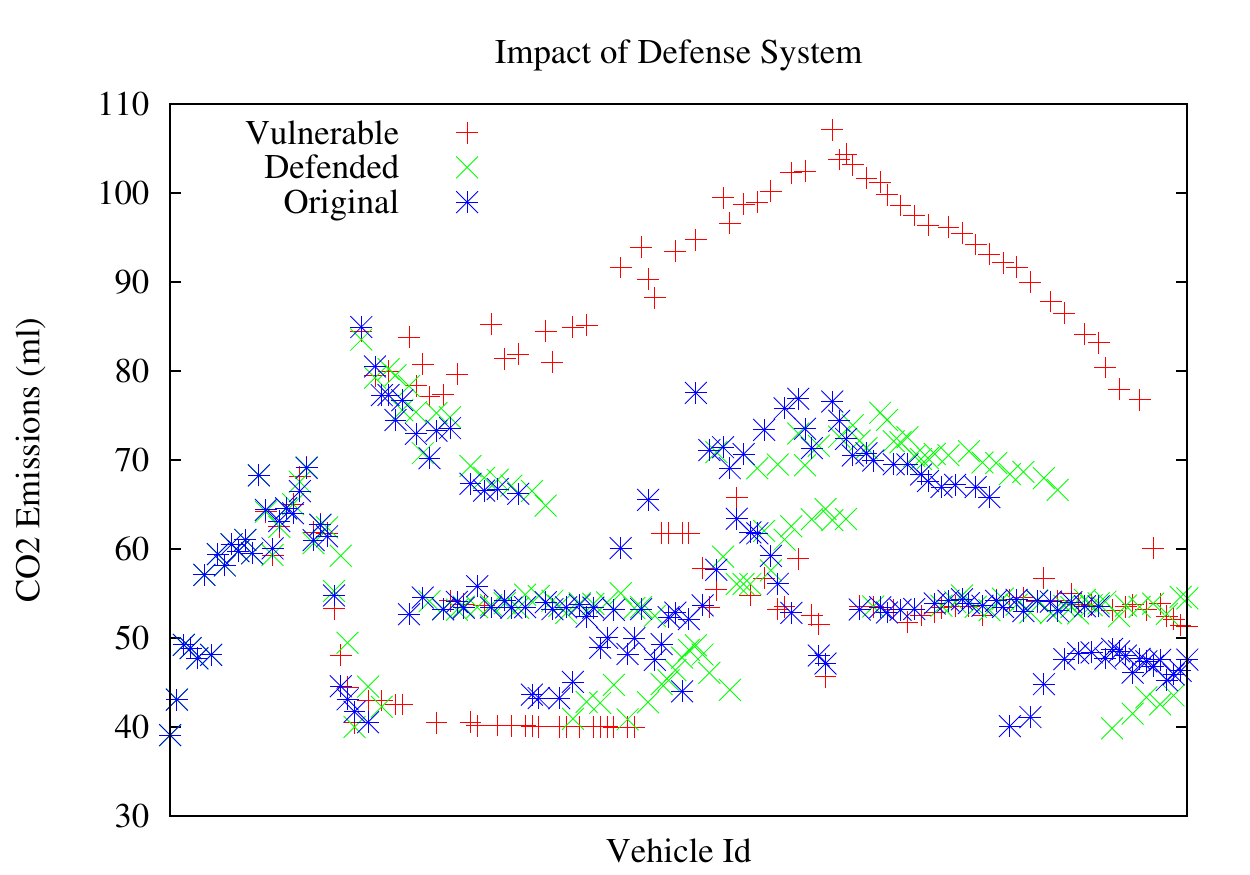}
    \includegraphics[width=\linewidth]{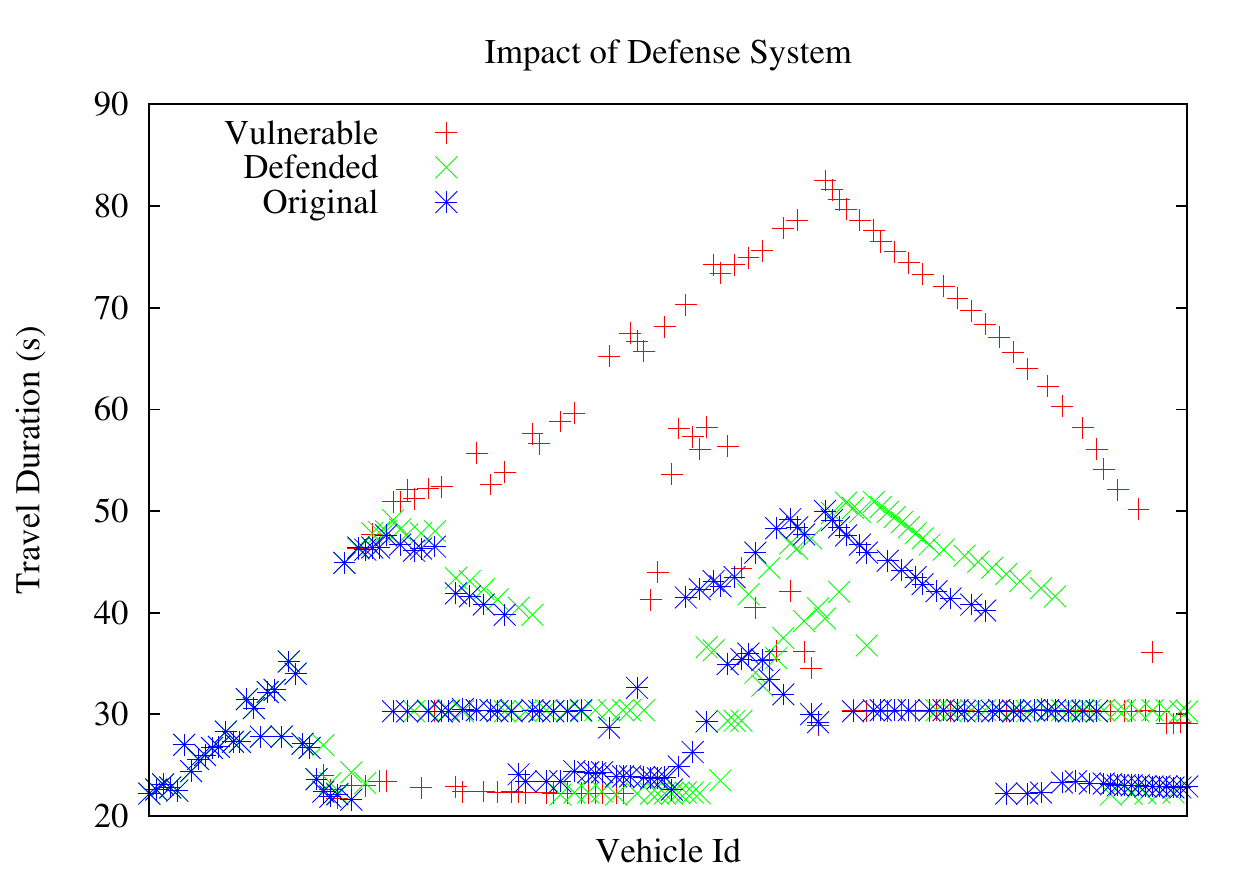}
  \caption{The proposed defense system  returns the protocol to near identical routing decisions by successfully filtering out the outliers and thus the overall system's performance is  preserved. }
  \label{impact-of-defense}
\end{figure}

In this last subsection, we present the performance of the proposed defense system against FD attacks. Recall that our goal is to have a performance similar to that of a scenario where no bogus data are running through the system and thus, prove the robustness of our defense mechanism.  Figure~\ref{impact-of-defense} illustrates the obtained results and it is evident
that the proposed method remarkably closely follows the performance of the original ErouVe algorithm. This is due to the fact that tweaked data are successfully omitted from the system and hence, ErouVe's routing instructions are only guided through real information.
The proportion of vehicles sent to the longer route is 26.5\% for the defended ErouVe and about 19\% for those that are vulnerable. 

The deviation observed between the defended and original algorithm can be explained by the following reasons: first, since tweaked data come in groups, i.e. three consecutive vehicles, when labeled bogus and thus omitted from the system, ErouVe is left with no new received reports for an interval between the last received bogus data and the most recent true report.  Second, a similar delay is induced in the protocol when data appears to be bogus, but it really is not, representing a traffic shift, between the time the report is labeled as BPS and later integrated in VoW. Such considerations induce a delay in the routing decisions and consequently, a deviation from the original ErouVe, but nevertheless are essential in order to filter out malicious vehicles.

\section{Conclusion}
In this paper we investigated how an eco-routing mechanism that is based on DSRC communications, 
is affected from faulty information that is  disseminated from malicious nodes in a vehicular environment. 
We implemented and tested the eco-routing mechanism
under attack scenarios that try to favor or discourage cars from following a route and we observed that a typical 
eco routing mechanism in an unprotected mode is strongly influenced by those attacks.
Based on these observations, we implemented novel defense mechanisms that exploit vehicular 
communications in order to make the network robust to several attacks. The defense mechanisms managed to alleviate
the effect of the attacks and restore the performance of the eco routing mechanism to near its optimal operation.
In the future, different attack scenarios are going to be investigated and more complex defense mechanisms developed.
The presented work can be a basis for the development of an integrated defense system for vehicular networks that 
can cope with  complex attack scenarios.

\bibliographystyle{IEEEtran}
\bibliography{bare_conf}

\end{document}